\documentclass[a4paper,10pt,twoside]{cpc-hepnp}

\usepackage{multicol}
\usepackage{graphicx}
\usepackage{booktabs}
\usepackage{amssymb,bm,mathrsfs,bbm,amscd}
\usepackage[tbtags]{amsmath}
\usepackage{lastpage}
\usepackage{CJK}
\usepackage{color}

\def\aap{Astron.\ Astrophys.}
\def\apj{Astrophys.\ J.}

\def\physrep{Phys.\ Rept.}
\def\prd{Phys.\ Rev.\ D}

\def\araa{Annu.\ Rev.\ Astron.\ Astrophys.}

\begin{document}
\begin{CJK*}{GBK}{song}


\fancyhead[c]{\small Submitted to Chinese Physics C} \fancyfoot[C]{\small 010201-\thepage}


\title{Reconciling the light component and all-particle cosmic ray energy spectra at the knee}

\author{%
      ZHAO Yi$^{1,2;1)}$\email{zhaoyi@ihep.ac.cn}%
\quad JIA Huan-Yu$^{1;2)}$\email{hyjia@home.swjtu.edu.cn}%
\quad ZHU Feng-Rong$^{1}$
}
\maketitle

\address{%
$^1$ Southwest Jiaotong University, Chengdu 610031, P.R.China\\
$^2$ Key Laboratory of Particle Astrophysics, Institute of High Energy Physics,
Chinese Academy of Science, Beijing 100049, P.R.China\\
}

\begin{abstract}
The knee phenomenon of the cosmic ray spectrum, which plays an important role in studying the acceleration mechanism of cosmic rays, is still an unsolved mystery. We try to reconcile the knee spectra measured by ARGO-YBJ
and Tibet-$\uppercase\expandafter{\romannumeral3}$. A simple broken
power-law model fails to explain the experimental data. Therefore
a modified broken power-law model with non-linear acceleration effects
is adopted, which can describe the sharp knee structure. This model
predicts that heavy elements dominate at the knee.
\end{abstract}




\begin{multicols}{2}

\section{Introduction}

The energy spectrum of cosmic rays, an important key to realize the
origin and acceleration mechanism of cosmic rays, can be simply
described by a power law over many magnitudes of energy from $10^{9}$ eV
to $10^{20}$ eV \citep{2009PrPNP..63..293B}, except for a few distinctive structures,
such as the knee phenomenon around $4 \times 10^{15}$ eV. The spectral
power index rapidly steepens from about $-2.7$ before the knee to $-3.1$
over the knee \citep{1958JETP...35...635K}. Numerous works on
 cosmic ray propagation and acceleration mechanisms
\citep{2009PrPNP..63..293B,1993A&A...268..726P,2002JHEP...12..033C,1999JETP...89..391B,
1993A&A...274..902S,1982ApJ...255..716J,1987ApJ...313..842J,
2002PhRvD..66h3004K,2004A&A...417..807V}
have been done to investigate the origin of the knee. These include the
diffusive shock acceleration (DSA) spectra origin of the knee based on the
theory of non-linear diffusive particle acceleration by shock waves from
supernova remnants (SNRs) \citep{2005JPhG...31R..95H,2012APh....35..801G}, and
the contribution of nearby pulsar wind producing very hard spectra in the
power-law spectra in the knee region \citep{2005ICRC....3..117B,2011CERNCourier...51...21E}.
However, the origin of the knee structure is still unconfirmed. Currently, the interpretations
of this phenomenon are based on phenomenological models from experimental measurements.
Many results have been given by direct observations with
balloon detectors \citep{2005ICRC....3..105W,2005ICRC....3..101S,2008ApJ...678..262A}
and indirect observations with air-shower detectors
on high altitude mountains \citep{2014ChPhC..38d5001B,2008ApJ...678.1165A,2004ApJ...612..268O}.
The CREAM experiment, as a balloon detector, has announced precise measurements of energy spectra
for individual nuclei ranging from $2$ TeV to $200$ TeV \citep{2011ApJ...728..122Y}.
As an air-shower ground-based detector located $4300$ m above sea level, the
Tibet-$\uppercase\expandafter{\romannumeral3}$ array has presented an all-particle
energy spectrum of primary cosmic rays from $100$ TeV to $100$ PeV, revealing a
sharp knee structure around $4$ PeV. Another air-shower experiment with
the same altitude as Tibet-$\uppercase\expandafter{\romannumeral3}$,
ARGO-YBJ, recently gave a new measurement of the energy spectrum of hydrogen and
helium nuclei, exhibiting a clear knee structure \citep{2015arXiv150203164B}.

The latest measurements released by ARGO-YBJ are investigated by the joint operation
of Resistive Plate Chambers (RPC) detectors and a Cherenkov telescope. This hybrid
detection bridges the gap between balloon detectors and ground-based experiments,
improves the shower energy resolution and enhances the capability to discriminate
showers induced by light nuclei from events initiated by heavier nuclei. They yield
clear evidence for a knee-like structure in the spectrum of hydrogen and helium nuclei
under 1 PeV \citep{2015arXiv150203164B}.

In the following discussions, we explain the knee spectra measured by ARGO-YBJ and
Tibet-$\uppercase\expandafter{\romannumeral3}$ by phenomenological models.
The superposition of energy spectrum of individual components described by a simple
broken power-law cannot reconcile the light components and all-particle energy spectra.
We therefore propose a parametric model to reconcile them. In Sec. $2$ we give two different
scenarios and parameters to fit the experimental data.
Sec. $3$ presents the average mass. Finally, Sec.$4$ gives some discussion and conclusions.

\section{Energy spectra formulation}

Cosmic rays emitted from the sources are most likely accelerated in the strong
shock fronts of SNRs by the DSA mechanism. The particles are deflected by chaotic magnetic fields, cross shock fronts frequently, and therefore gain energy up to the PeV region.
This acceleration leads to the observed approximate power-law spectrum. The energy
spectrum is modified during diffusive propagation. The spectra near the sources
vary from that observed on the earth, which may due to nuclear spallation or decay,
ionization losses, leakage from the galaxy and solar modulation for low energies
\citep{1987PhR...154....1B,1980ARA&A..18..289C,1993ApJ...414..601L}. We try to
adopt two different phenomenological models to explain the observed energy spectra
around knee structure. One is an expression of a simple broken power-law and the other is
the broken power-law with non-linear modification term.

\subsection{Broken power-law model and parameters}
The all-particle primary energy spectra derived from extensive air shower experiments
can be described by a broken power-law function. We therefore use this kind of function
to parameterize the observed differential energy spectra for individual
elements of cosmic ray, which can be written as
\begin{eqnarray}\label{eq:zy1}
\frac{dj}{dE} = j_{0}E^{-\gamma}(1+\frac{E}{z\epsilon_{b}})^{-\Delta\gamma} ,
\end{eqnarray}
where $\frac{dj}{dE}$ is the differential flux of the individual element with energy $E$,
$j_{0}$ is the normalization constant, $\epsilon_{b}$ represents the break point of the proton spectrum,
$z$ is the atomic number, $\gamma$ is the power index in the energy range $E<<z\epsilon_{b}$,
and $\gamma+\Delta\gamma$ denotes the power index when $E>>z\epsilon_{b}$.
We try to use this expression to fit both the combined energy spectrum of hydrogen and helium nuclei
observed by ARGO-YBJ and the all-particle energy spectrum given by Tibet-$\uppercase\expandafter{\romannumeral3}$.
The interaction model SIBYLL for Tibet data is used in this work. We firstly fix
the energy spectra of each component below the knee region, and then extrapolate
the spectra beyond the knee. The observed data of individual elements ($H$, $He$, $C$, $O$, $Ne$, $Mg$, $Si$ and $Fe$)
from CREAM are adopted. For the spectrum index of each element, see \citep{2011ApJ...728..122Y,2009ApJ...707..593A}.
An expression of simple power-law $dj/dE=jE^{-\gamma}$ is used to fit the spectrum of
each component, with the fitting parameters $j$ and spectral indices listed in Table 1.

\begin{center}
\tabcaption{Constant $j$ and spectrum index $\gamma$ below the knee region.}
 \begin{tabular}{cccc}
  \hline\noalign{\smallskip}
   $z$      & Element  & $j(eV^{-1}m^{-2}s^{-1}sr^{-1})$  & $\gamma$ \\
  \hline\noalign{\smallskip}
  1             & H         & $6.62\times10^{18}$         &2.66  \\
  2             & He        & $6.40\times10^{17}$         &2.58  \\
  6             & C         & $2.43\times10^{17}$         &2.61  \\
  8             & O         & $1.88\times10^{18}$         &2.67  \\
  10            & Ne        & $1.68\times10^{18}$         &2.72  \\
  12            & Mg        & $4.56\times10^{17}$         &2.66  \\
  14            & Si        & $7.99\times10^{17}$         &2.67  \\
  26            & Fe        & $6.71\times10^{17}$         &2.63  \\
  \noalign{\smallskip}\hline
\end{tabular}
\end{center}

With Equation (\ref{eq:zy1}), we fit the combined spectrum of hydrogen and helium nuclei
observed by ARGO-YBJ \citep{2012PhRvD..85i2005B,2014ChPhC..38d5001B,2015arXiv150203164B}, and get the
fitting result of energy break point $\epsilon_{b}=4\times10^{14}eV$ and $\Delta\gamma=0.68$
for the proton. The break point for other elements can be calculated by $z\epsilon_{b}$.
The spectral indices of all elements at the higher energy range beyond the knee region are assumed to be
the same value as the result from ARGO-YBJ, i.e. $-\gamma-\Delta\gamma=-3.34$. We sum up the energy spectra
of all elements to get the approximate all-particle spectrum, seen as the red solid line in
Fig. \ref{fig:zy1}. It can be seen from Fig. \ref{fig:zy1} that this all-particle spectrum
based on ARGO-YBJ observation with this simple broken power-law model cannot match the
observed all-particle spectrum given by Tibet-$\uppercase\expandafter{\romannumeral3}$.
The simple broken power-law function therefore cannot describe the energy spectrum in the
knee region, which perhaps is due to the influence of the DSA mechanism.

\begin{center}
\includegraphics[width=0.9\columnwidth, angle=0]{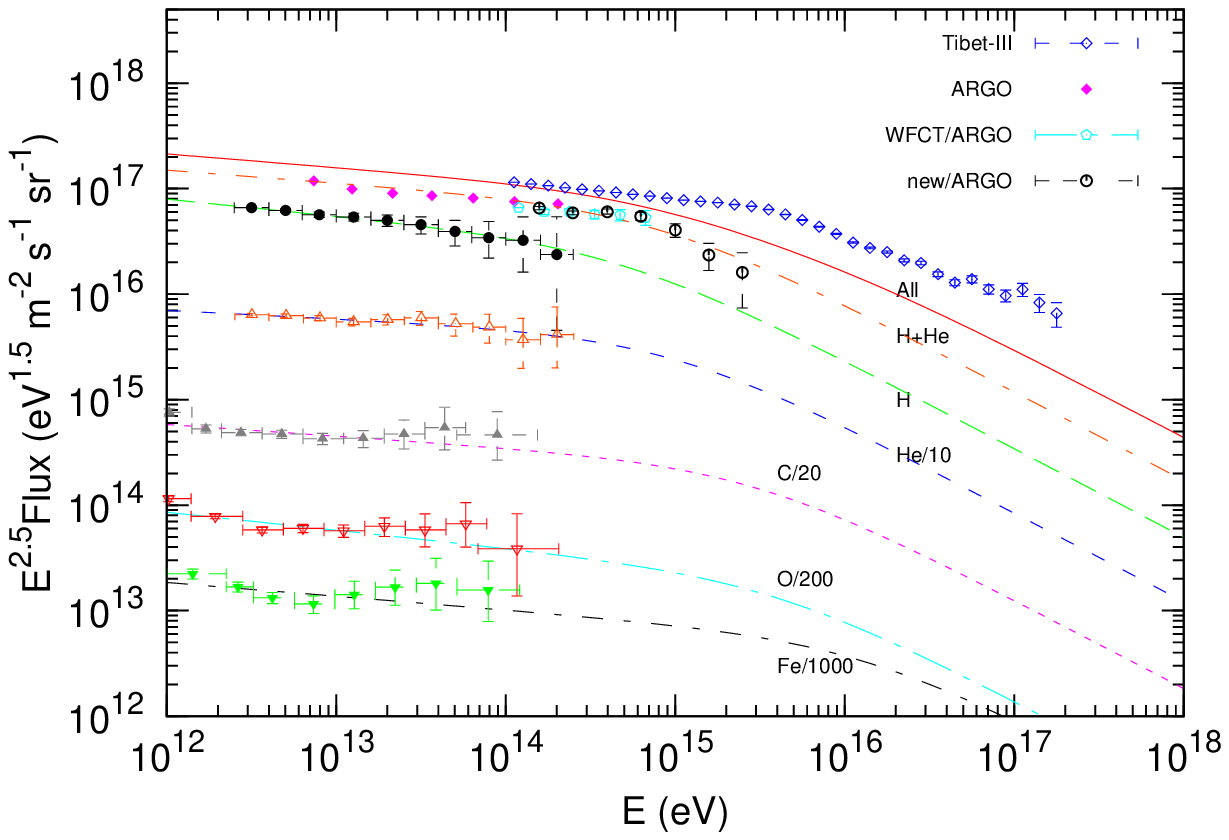}
\figcaption{All-particle spectrum (red solid line) calculated as the sum of individual components, with spectra
of some individual elements. The data are cited from CREAM \citep{2011ApJ...728..122Y,2009ApJ...707..593A},
ARGO-YBJ \citep{2012PhRvD..85i2005B,2014ChPhC..38d5001B,2015arXiv150203164B} and
Tibet-$\uppercase\expandafter{\romannumeral3}$ \citep{2008ApJ...678.1165A}.}
\label{fig:zy1}
\end{center}

\subsection{Non-linear model and parameters}
Since the differential energy spectrum in the form of the simple broken power-law cannot reconcile
the observations of ARGO-YBJ and Tibet-$\uppercase\expandafter{\romannumeral3}$,
a new correction term is added, so the broken power-law can now be expressed as
\begin{eqnarray}\label{eq:zy2}
\frac{dj}{dE} = j_{0}E^{-\gamma}[1+\alpha(\frac{E}{z\epsilon_{b}})^{\beta}](1+\frac{E}{z\epsilon_{b}})^{-\Delta\gamma} ,
\end{eqnarray}
where $\alpha$ is assumed to be a $z$-dependent parameter, and $\beta$ is a constant.
The DSA mechanism may result in a spectrum with concave-up curvature,
so we multiply the non-linear term $1+\alpha(\frac{E}{z\epsilon_{b}})^{\beta}$ into the
simple broken power-law to structure a concave-up form as a phenomenological model.
Without this non-linear term, the superimposed all-particle spectrum from all elements
is smooth and cannot fit the sharp knee structure as measured by  Tibet-$\uppercase\expandafter{\romannumeral3}$. In addition, we have tried some parameters in a non-$z$-dependent form, but these cannot
match the data, so we finally adopt the form in Equation (\ref{eq:zy2}).
An equation to express the non-linear effects is proposed in \citep{2010ApJ...716.1076S}.
They propose the non-linear process by a modified exponential cut-off
expression for the spectra from sources, and then deduce the observed spectra
by the superposition of these spectra. From their formula, the model they deduced is
a kind of broken power-law multiplied by a complex modified term. While our expression form
is simpler, we modify the simple broken power-law directly as the observed spectra and
then parameterize this scenario. Combined with the newly published measurements from ARGO-YBJ,
our model is easier to quantify the parameters precisely. A model with precisely confirmed
parameters will help us to review its intrinsic physical issues.
From Equation (\ref{eq:zy2}), we get a good fit result to match the observations
from both ARGO-YBJ and Tibet-$\uppercase\expandafter{\romannumeral3}$ with parameters
$\alpha=1.50+0.35z$, $\beta=1.20$, $\epsilon_{b}=2.0\times10^{14}eV$ and $\Delta\gamma=1.88$
for the proton. $\Delta\gamma$ values of other elements are assumed to obey the formula
$-\gamma-\Delta\gamma+\beta=-3.34$. The fitting spectra are plotted in Fig. \ref{fig:zy2},
where the solid red  line represents the all-particle spectrum. From Fig. \ref{fig:zy2}
we can see that the all-particle spectrum based on ARGO-YBJ observation can match the
observed all-particle spectrum given by Tibet-$\uppercase\expandafter{\romannumeral3}$.
Based on this model, the distinctive knee structure is mostly contributed by heavy elements
such as $Fe$.

\begin{center}
\includegraphics[width=0.9\columnwidth, angle=0]{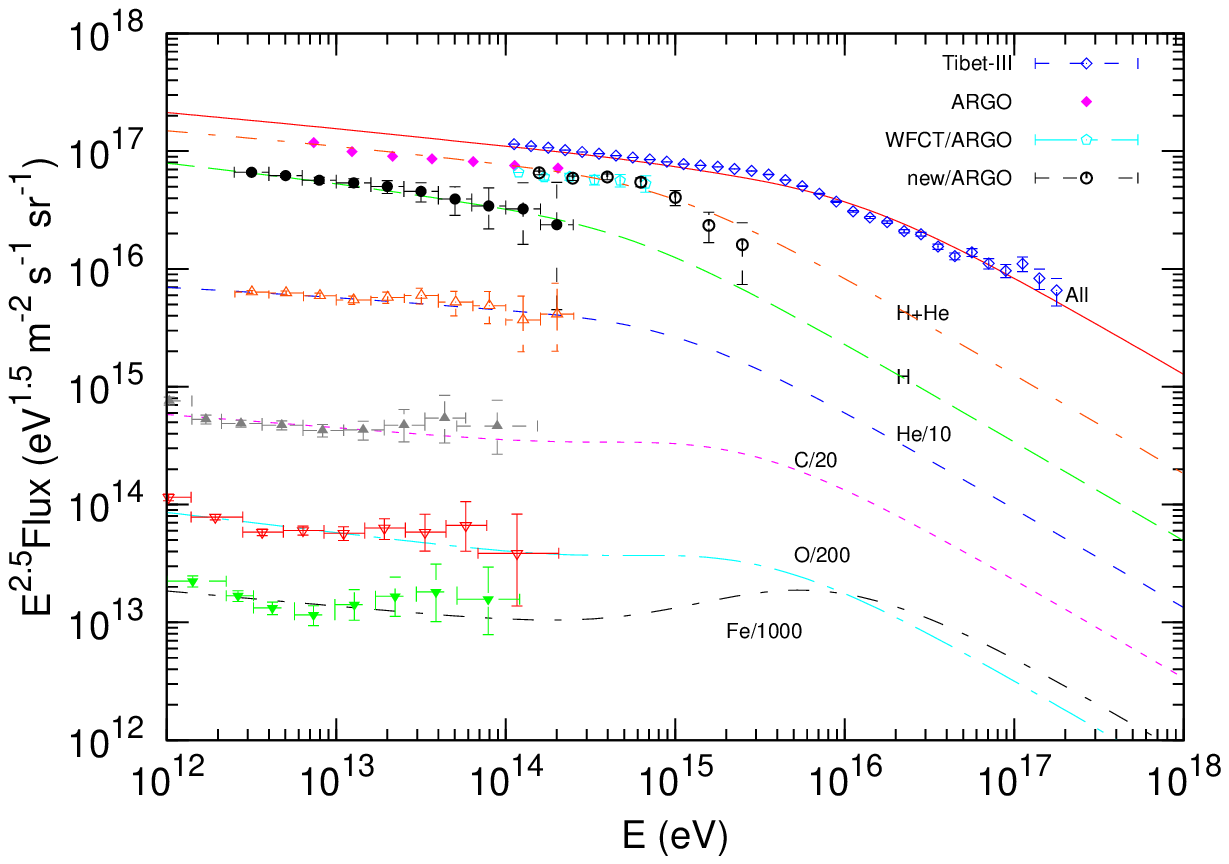}
\figcaption{All-particle spectrum (solid red line) calculated as the sum of individual components, with spectra
of some individual elements. The data are cited from CREAM \citep{2011ApJ...728..122Y,2009ApJ...707..593A},
ARGO-YBJ \citep{2012PhRvD..85i2005B,2014ChPhC..38d5001B,2015arXiv150203164B} and
Tibet-$\uppercase\expandafter{\romannumeral3}$ \citep{2008ApJ...678.1165A}.}
\label{fig:zy2}
\end{center}

In order to obtain the best fitting parameters for the non-linear model, we employ the
Markov Chain Monte Carlo (MCMC) \citep{2002PhRvD..66j3511L,2012PhRvD..85d3507L},
a very efficient algorithm, to achieve our goal.
Considering that the calculation of $\chi^{2}$ will be affected by the too small errors
of all-particle spectrum given by Tibet-$\uppercase\expandafter{\romannumeral3}$,
we amplify the errors of this spectrum by taking account of energy resolution.
The initial form of data with errors listed in the literature \citep{2008ApJ...678.1165A}
are expressed by $F \pm \Delta F$, where $F$ denotes the flux and
$\Delta F$ denotes the systematic error. As a new deduced additional error,
a new term is added to this expression, i.e. $F \pm \Delta F \pm F(\gamma-1)(\Delta E/E)$,
where $\gamma$ denotes the spectrum index. We take the energy resolution $\Delta E/E =0.15$
as a constant. The following $5$ parameters: $\alpha=p_{1}+p_{2}z$, $\beta$,
$\epsilon_{b}$ and $\Delta\gamma$, are set free for the MCMC algorithm, where
$\alpha$ consists of free parameters $p_{1}$, $p_{2}$ and redshift $z$.
The parameters $\epsilon_{b}$ and $\Delta\gamma$ here refer to protons as mentioned above.
Because every spectrum index of elements in the higher energy band beyond the knee is supposed
to be the same value, we can get the $\Delta\gamma$ of other components from the
$\Delta\gamma$ of the proton. By this MCMC algorithm, we obtain the best fitting parameters as follows:
$\alpha=2.21+0.45z$, $\beta=1.62$, $\epsilon_{b}=1.0\times10^{14}eV$ and $\Delta\gamma=2.25$
for the proton. The parameters $\alpha$, $\beta$ and $\epsilon_{b}$ decide the shapes of the energy spectra.
Different values result in different spectra, as shown in Fig. \ref{fig:zy3},
where the elements $H$ and $Fe$ are taken as examples.

\begin{center}
\includegraphics[width=0.48\columnwidth, angle=0]{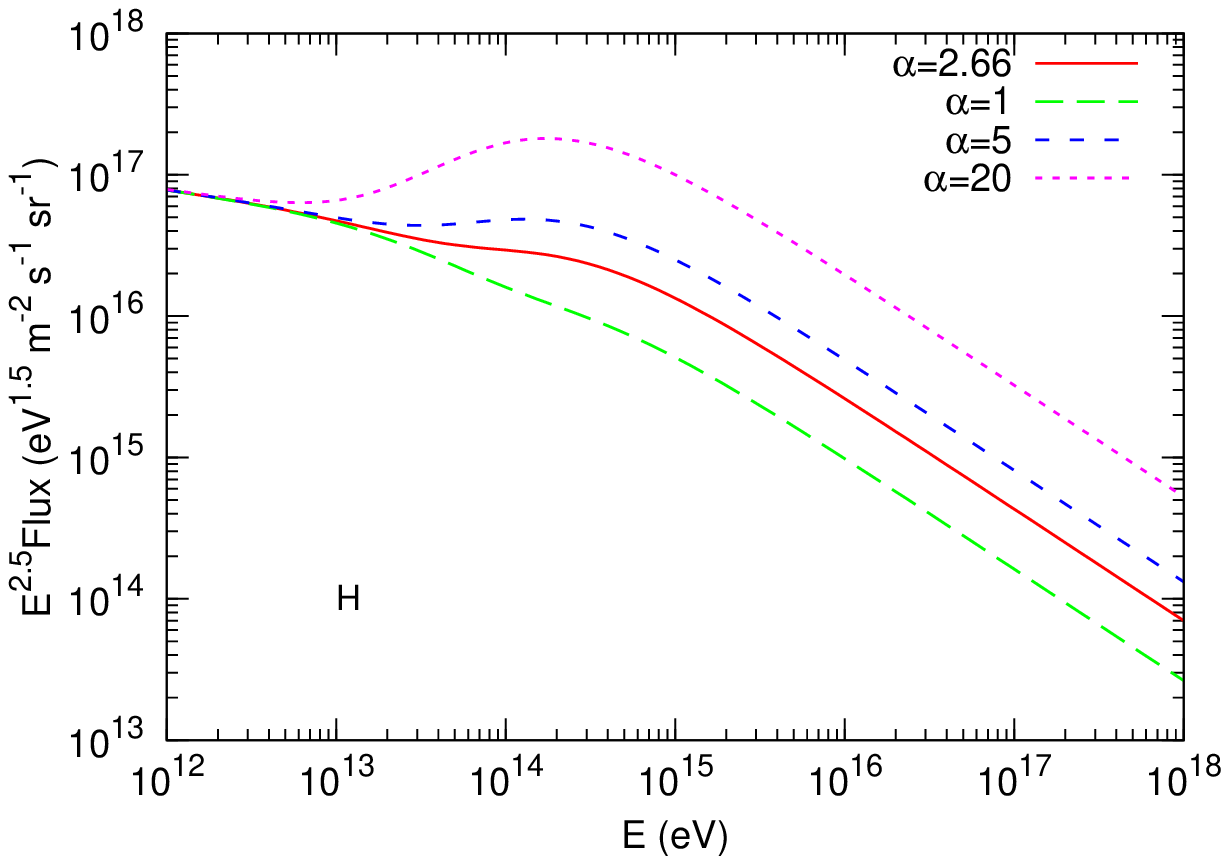}
\includegraphics[width=0.48\columnwidth, angle=0]{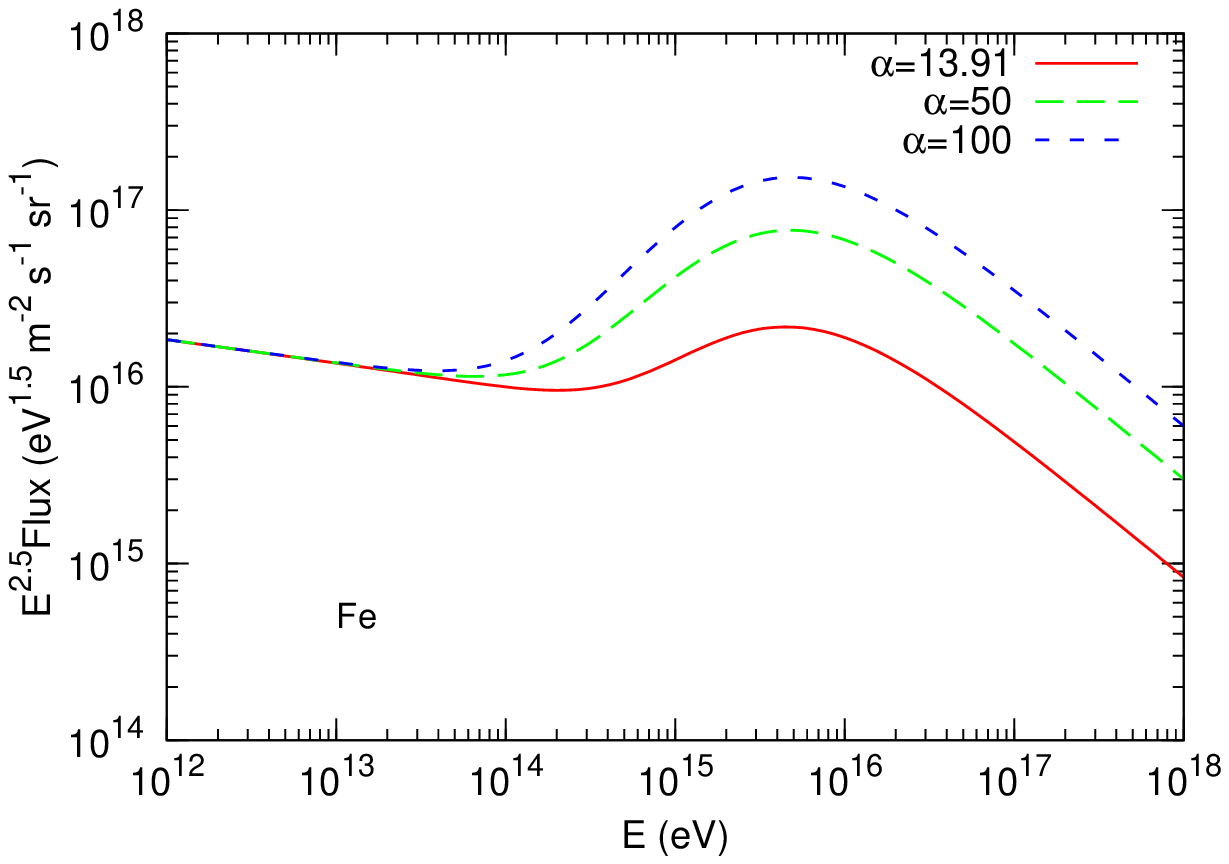}
\includegraphics[width=0.48\columnwidth, angle=0]{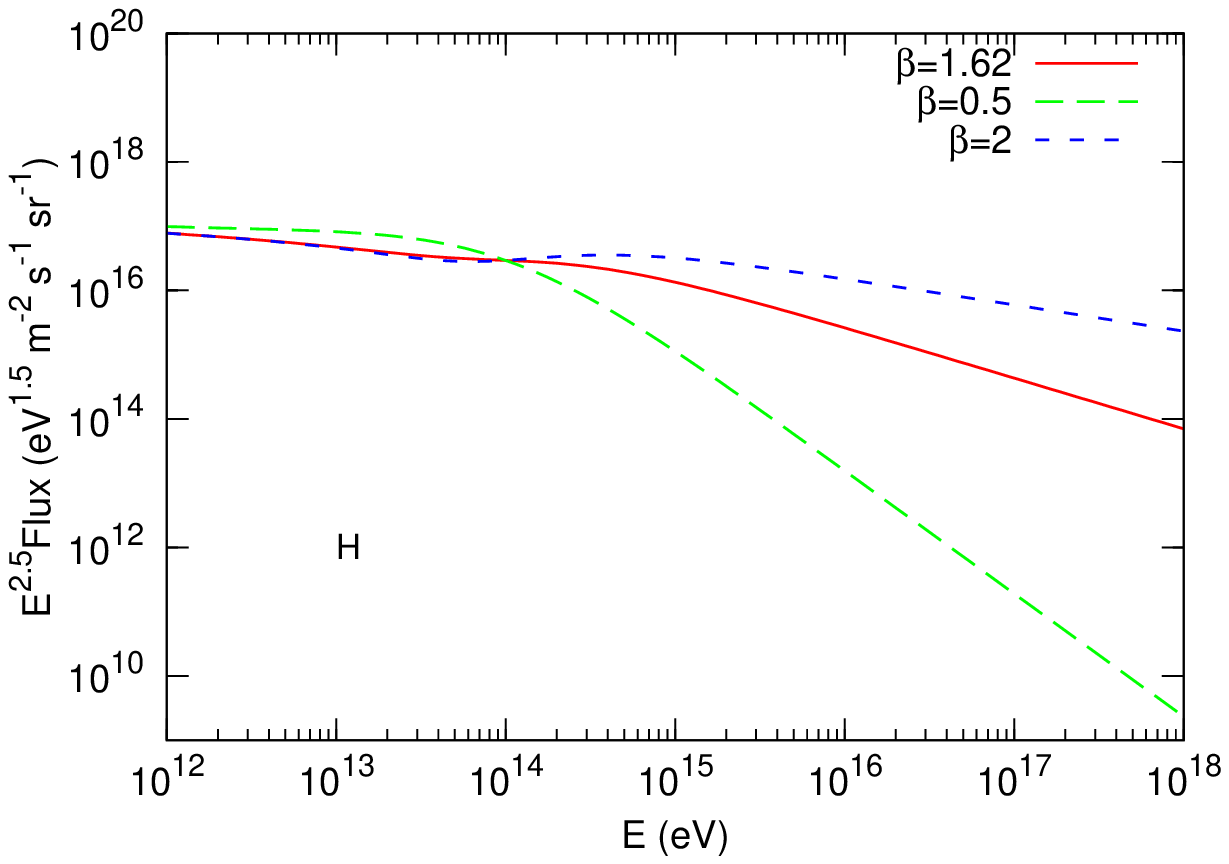}
\includegraphics[width=0.48\columnwidth, angle=0]{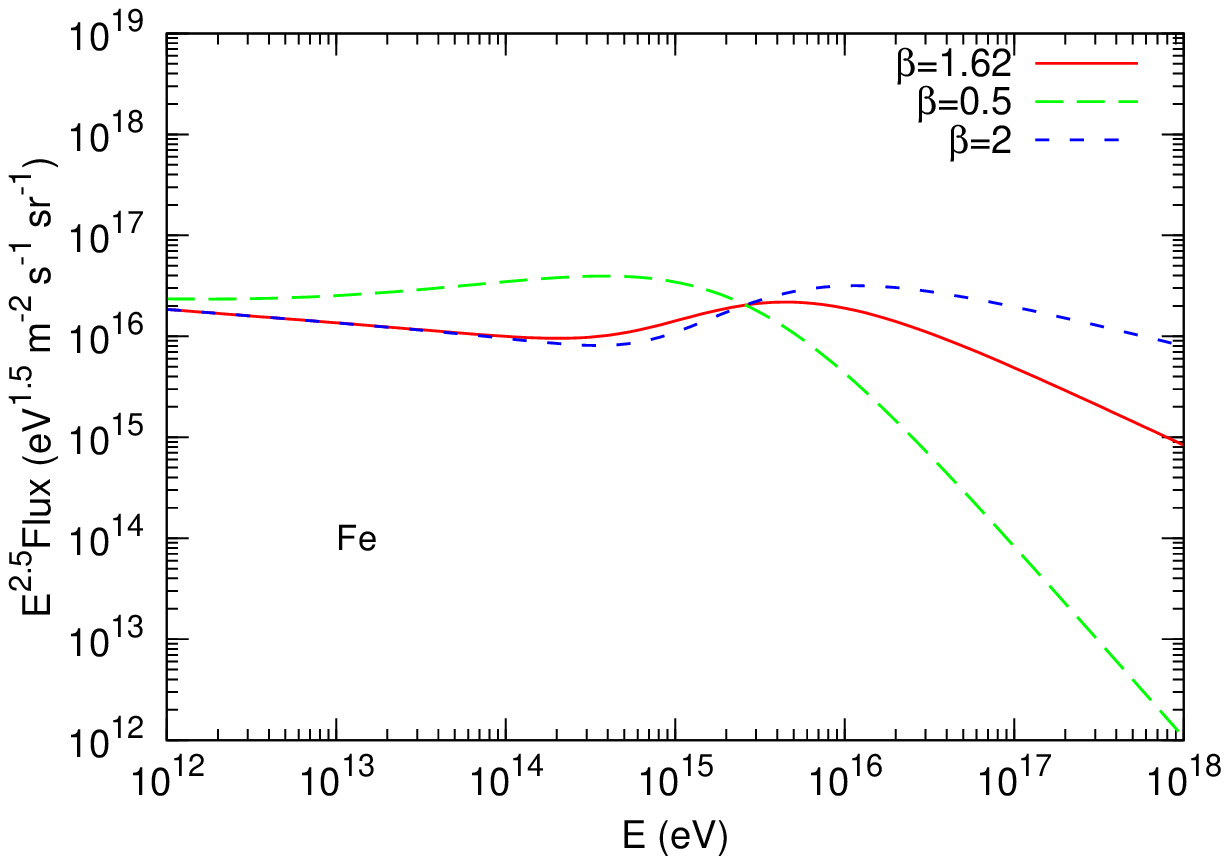}
\includegraphics[width=0.48\columnwidth, angle=0]{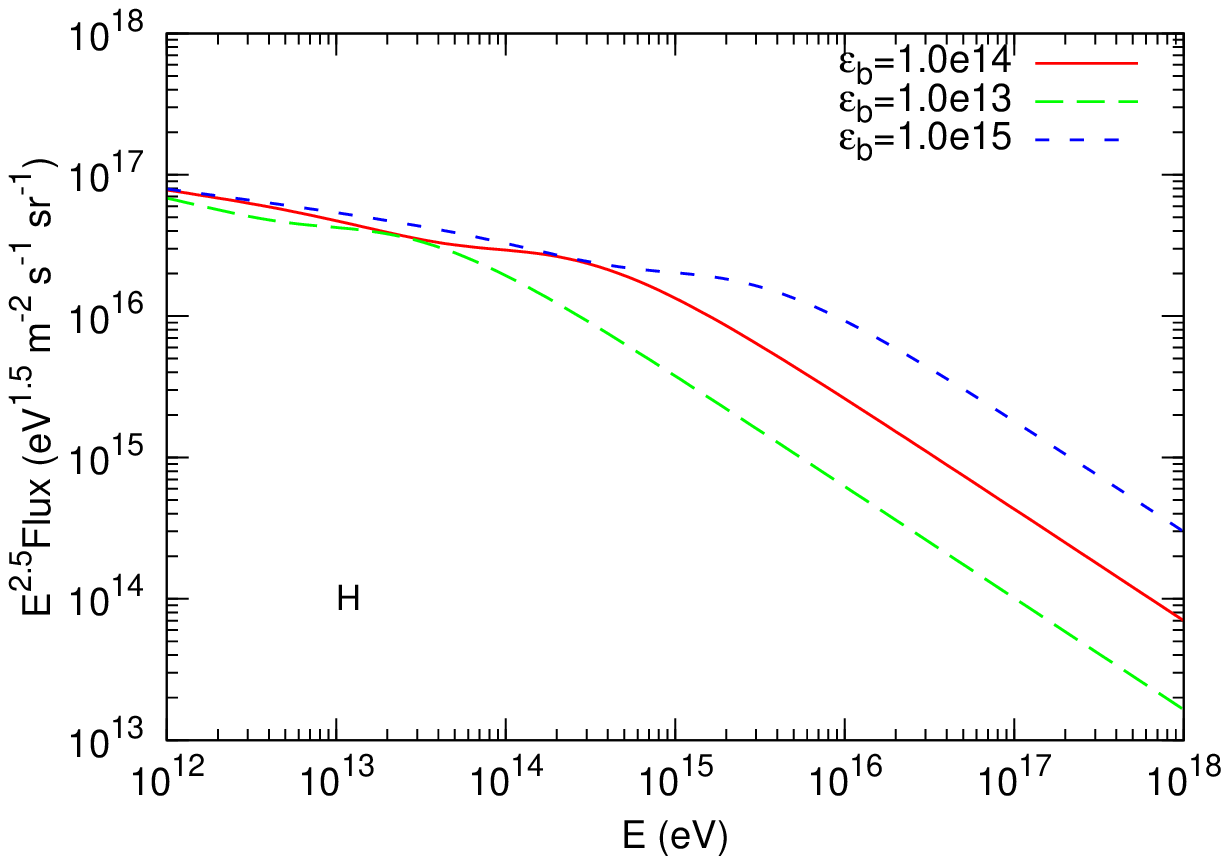}
\includegraphics[width=0.48\columnwidth, angle=0]{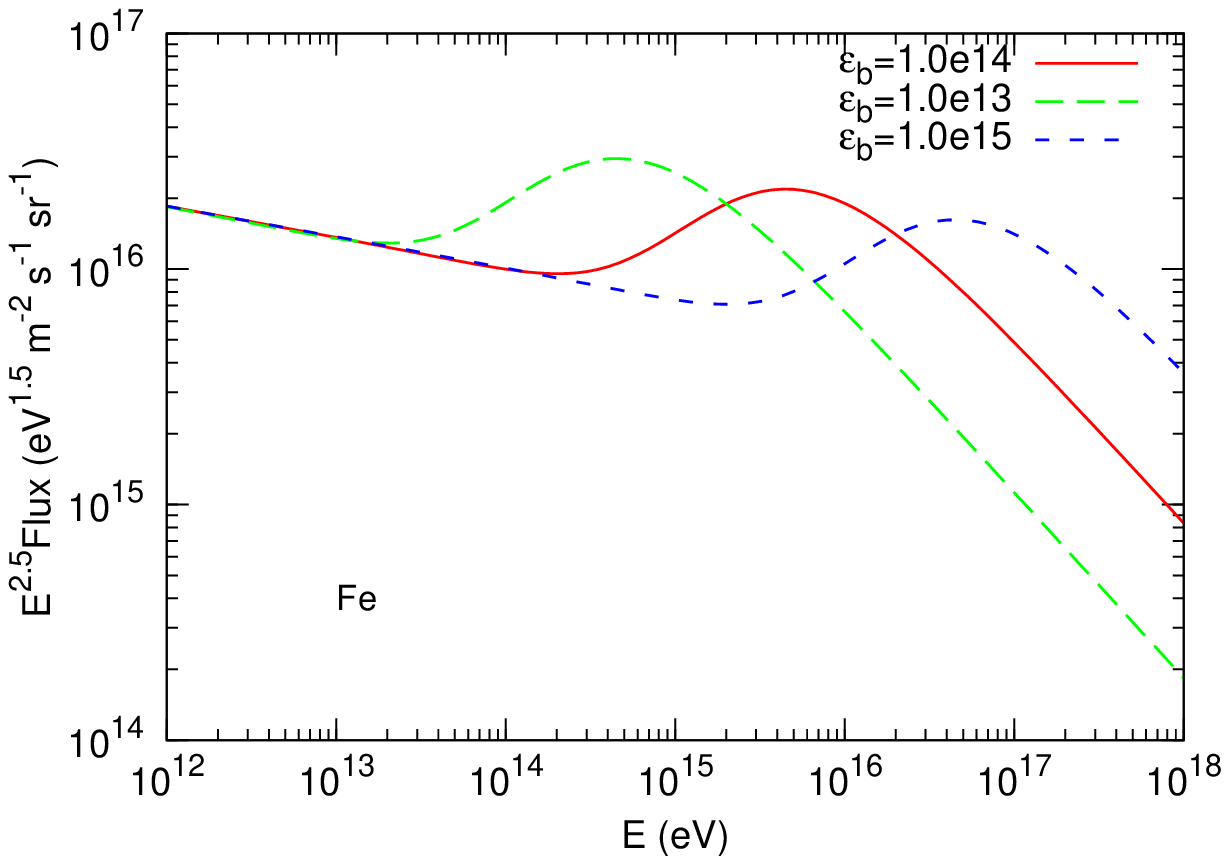}
\figcaption{Energy spectrum shapes determined by $\alpha$,
$\beta$ and $\epsilon_{b}$ for elements  $H$ (left) and $Fe$ (right).
The upper two figures represent the dependence of the spectra on $\alpha$,
the middle two figures represent the dependence on $\beta$,
and the lower two figures represent the dependence on $\epsilon_{b}$.
The solid red lines in all six figures represent the currently used values.}
\label{fig:zy3}
\end{center}

The best fit result is shown in Fig. \ref{fig:zy4}. The measurements of hydrogen and
helium nuclei spectra using the SIBYLL model by the KASCADE group \citep{2005APh....24....1A}
are compared with our model in Fig. \ref{fig:zy4}. The KASKADE group claimed that the energy spectra of both
proton and helium show a knee-like feature, but in our model, this seems not to be the case.

\begin{center}
\includegraphics[width=0.9\columnwidth, angle=0]{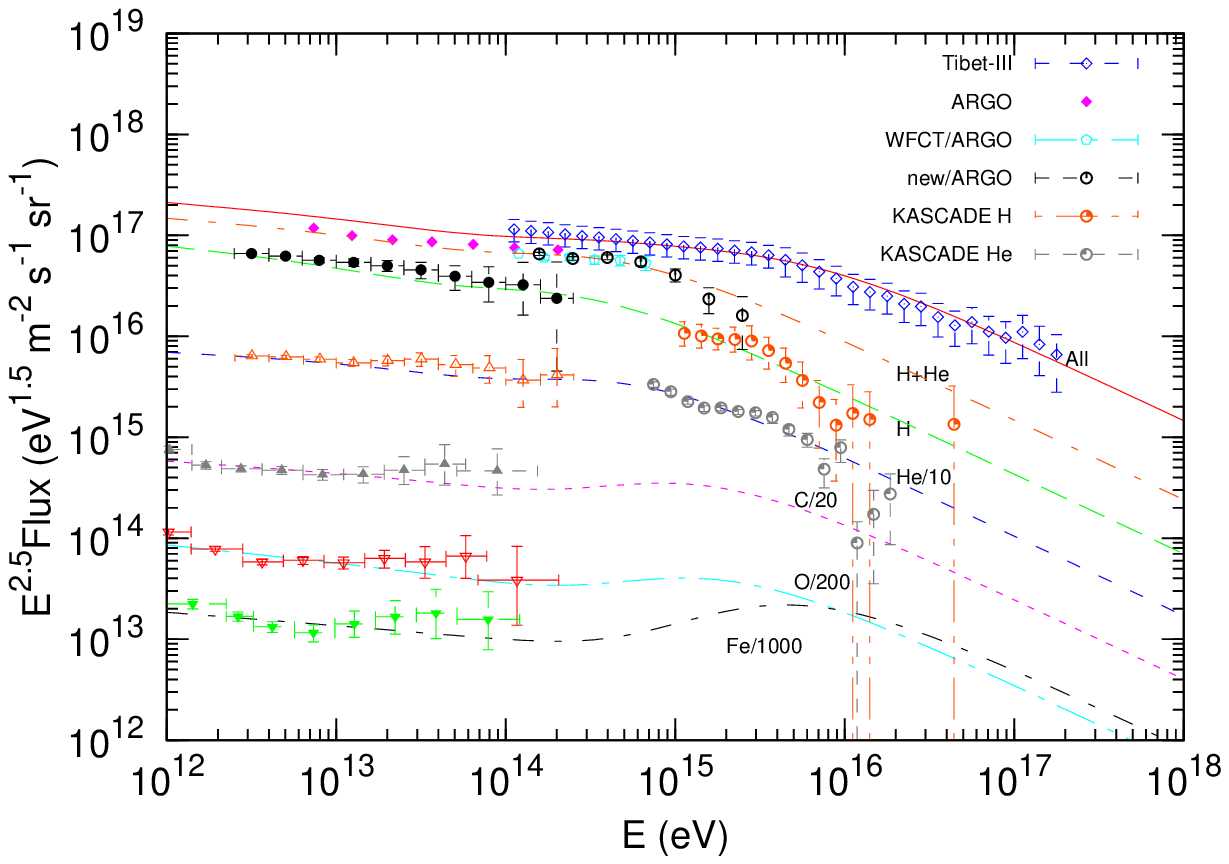}
\figcaption{All-particle spectrum (solid red line) calculated as the sum of individual components, with spectra
of some individual elements. The data are cited from CREAM \citep{2011ApJ...728..122Y,2009ApJ...707..593A},
ARGO-YBJ \citep{2012PhRvD..85i2005B,2014ChPhC..38d5001B,2015arXiv150203164B},
Tibet-$\uppercase\expandafter{\romannumeral3}$ \citep{2008ApJ...678.1165A} and KASCADE \citep{2005APh....24....1A}.}
\label{fig:zy4}
\end{center}

\section{Average mass}
A commonly-used quantity to characterize the mass composition of cosmic rays
is the mean logarithmic mass, which is defined as
\begin{eqnarray}
\langle lnA \rangle=\sum_{i} r_{i} ln A_{i},
\end{eqnarray}
$r_{i}$ being the relative fraction of nuclei of mass $A_{i}$.
The average mass is reported by many air shower experiments,
but the results are divergent because of the poor primary mass
resolution. We calculate the average mass for two cases:
one for the former non-linear model and another for the latter
best fitting case, as shown in Fig. \ref{fig:zy5}.

From our derived average mass of cosmic rays, we can see $\langle lnA \rangle$ goes up
between several hundreds of TeV and several tens of PeV. It implies that
heavier components occupy an increasing proportion with the increase of energy,
and predicts an iron-dominant composition above the knee. But the components
above the knee are not all $Fe$ , because $\langle lnA \rangle$ for $Fe$ is about 4.0.
Further experimental measurements of the chemical composition of high energy
cosmic rays will verify our prediction. If the light components dominate the
knee structure, this model can be excluded, while if the heavy components dominate
the knee structure, this model will be confirmed, and the measurements will be
helpful to improve the model parameters.

\begin{center}
\includegraphics[width=0.9\columnwidth, angle=0]{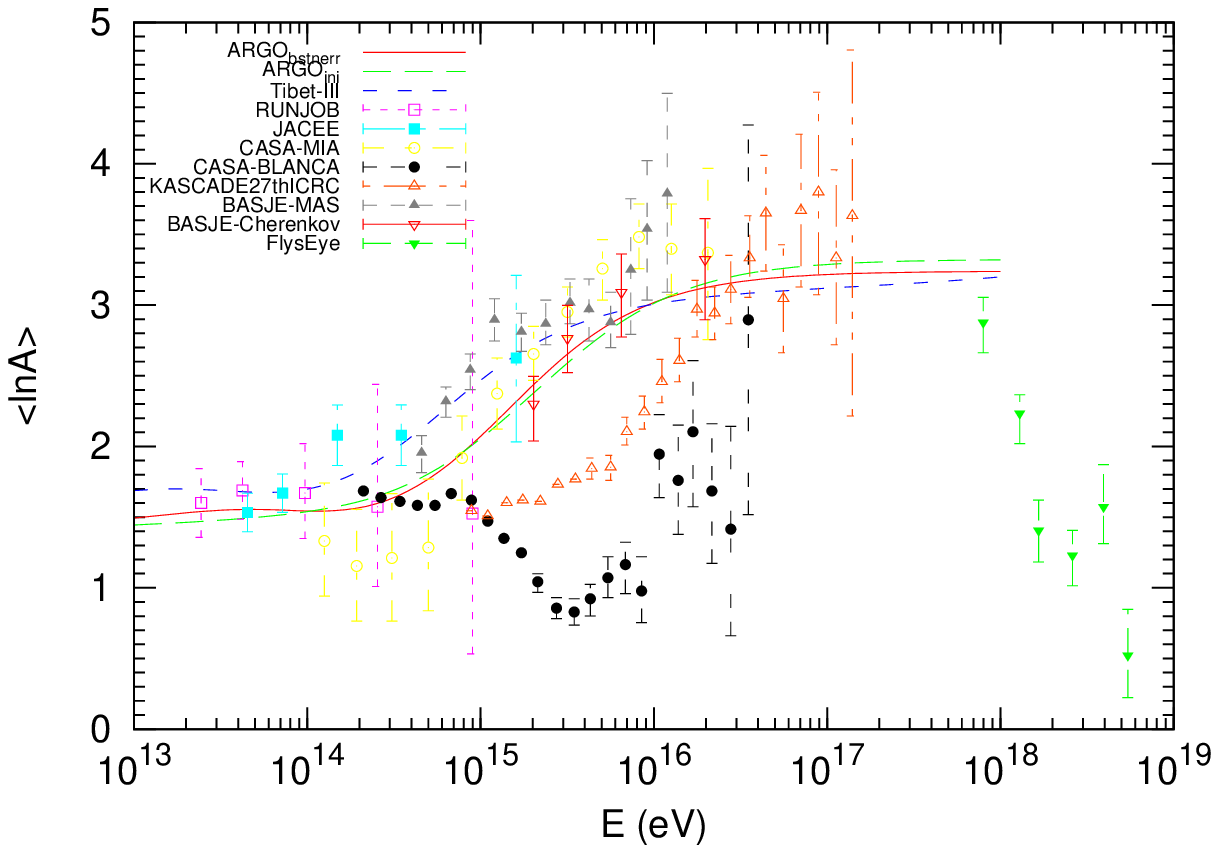}
\figcaption{Average mass calculated for the non-linear model, compared with that of
Tibet-$\uppercase\expandafter{\romannumeral3}$ (blue dashed line). The green
dashed line represents the first fitting result by the non-linear model and the
red solid line represents the best fitting result. The data are cited from
RUNJOB \citep{2001APh....16...13A}, JACEE \citep{1998ApJ...502..278A},
CASA/MIA \citep{1999APh....10..291G}, CASA/BLANCA \citep{2001APh....15...49F},
KASCADE 27thICRC \citep{2001ICRC....1...71H}, BASJE-MAS and BASJE-Cherenkov
\citep{2004ApJ...612..268O}, and Fly's Eye \citep{1993PhRvL..71.3401B}.}
\label{fig:zy5}
\end{center}

\section{Discussion and conclusion}
Many mechanisms have been discussed to explain the knee structure of
cosmic rays, including the diffusive shock acceleration mechanism of cosmic
rays in supernova explosion shock fronts, leakage from the galaxy, interactions
with background particles in the interstellar medium, and new physics of
interactions in the atmosphere. Some models are likely to be excluded, however,
such as the leaky box model, interactions with background particles and new types
of interactions in the atmosphere. The leakage from the galaxy does not give a distinct change of the spectral slope at the knee. The scenario of interactions with background particles yields a very light mean logarithmic mass, while the measurements indicate
an increase of mass with energy. For new types of interactions in the atmosphere,
there are no inconsistencies between the different air shower components by the
simultaneous observations from KASCADE, and the measurements of different air shower
components can be interpreted with standard particle physics \citep{2004APh....21..241H}.

We propose two scenarios to reconcile the light components and
all-particle energy spectra by phenomenological models.
In the first scenario, we use a simple broken power-law model to fit the
combined energy spectrum of hydrogen and helium nuclei measured by ARGO-YBJ
and the all-particle spectrum measured by Tibet-$\uppercase\expandafter{\romannumeral3}$,
but the all-particle spectrum cannot be fitted by this model. Hence, the simple
broken power-law fails to explain the knee structure. In the second scenario,
a non-linear modification term is added to the broken power-law. This modified broken
power-law model with non-linear acceleration effects can describe the sharp knee
structure. The mechanism of non-linear effects at shock fronts of SNRs has been
explained in earlier works \citep{2001RPPh...64..429M}. Particles with higher energies
can scatter farther ahead of the shock, which results in an effectively higher compression
ratio, so a more locally hard spectrum is produced, and a concave-up curvature to the
accelerated-particle distribution is expected \citep{2008ARA&A..46...89R}.
In addition, this modified model predicts that heavy elements dominate
at the knee. Based on this model, some models holding the viewpoints of weak
knee structure or protons dominating the knee are likely to be excluded. Precise measurements
of energy spectra of individual elements in the knee region will test our model.

\acknowledgments{This work is supported in part by the National Natural Science
Foundation of China (No. 11175147). We are grateful to Bi Xiaojun, Yuan Qiang
and Lin Sujie for helpful discussions.}

\end{multicols}

\vspace{10mm}

\begin{multicols}{2}

\end{multicols}

\clearpage

\end{CJK*}

\begin{thebibliography}{90}

\vspace{3mm}

\bibitem[Bl{\"u}mer et al.(2009)]{2009PrPNP..63..293B} Bl{\"u}mer, J., Engel, R., H{\"o}randel, J.~R.\ 2009, Progress in Particle and Nuclear Physics, 63, 293

\bibitem[Kulikov \& Khristiansen(1958)]{1958JETP...35...635K} Kulikov, G. V., \& Khristiansen, G. B.\ 1958, JETP, 35, 635

\bibitem[Ptuskin et al.(1993)]{1993A&A...268..726P} Ptuskin, V.~S., Rogovaya, S.~I., Zirakashvili, V.~N., et al.\ 1993, \aap, 268, 726

\bibitem[Candia et al.(2002)]{2002JHEP...12..033C} Candia, J., Roulet, E., \& Epele, L.~N.\ 2002, Journal of High Energy Physics, 12, 33

\bibitem[Berezhko \& Ksenofontov(1999)]{1999JETP...89..391B} Berezhko, E.~G., \& Ksenofontov, L.~T.\ 1999, Soviet Journal of Experimental and Theoretical Physics, 89, 391

\bibitem[Stanev et al.(1993)]{1993A&A...274..902S} Stanev, T., Biermann, P.~L., \& Gaisser, T.~K.\ 1993, \aap, 274, 902

\bibitem[Jokipii(1982)]{1982ApJ...255..716J} Jokipii, J.~R.\ 1982, \apj, 255, 716

\bibitem[Jokipii(1987)]{1987ApJ...313..842J} Jokipii, J.~R.\ 1987, \apj, 313, 842

\bibitem[Kobayakawa et al.(2002)]{2002PhRvD..66h3004K} Kobayakawa, K., Honda, Y.~S., \& Samura, T.\ 2002, \prd, 66, 083004

\bibitem[V{\"o}lk \& Zirakashvili(2004)]{2004A&A...417..807V} V{\"o}lk, H.~J., \& Zirakashvili, V.~N.\ 2004, \aap, 417, 807

\bibitem[Hillas(2005)]{2005JPhG...31R..95H} Hillas, A.~M.\ 2005, Journal of Physics G Nuclear Physics, 31, 95

\bibitem[Gaisser(2012)]{2012APh....35..801G} Gaisser, T.~K.\ 2012, Astroparticle Physics, 35, 801

\bibitem[Bhadra(2005)]{2005ICRC....3..117B} Bhadra, A.\ 2005, International Cosmic Ray Conference, 3, 117

\bibitem[Erlykin et al.(2011)]{2011CERNCourier...51...21E} Erlykin, A., Martirosov, R., \& Wolfendale, A.,\ 2011, CERN Courier, 51(1), 21

\bibitem[Wefel et al.(2005)]{2005ICRC....3..105W} Wefel, J.~P., Adams, J.~H., Ahn, H.~S., et al.\ 2005, International Cosmic Ray Conference, 3, 105

\bibitem[Seo et al.(2005)]{2005ICRC....3..101S} Seo, E.~S., Ahn, H.~S., Allison, P., et al.\ 2005, International Cosmic Ray Conference, 3, 101

\bibitem[Ave et al.(2008)]{2008ApJ...678..262A} Ave, M., Boyle, P.~J., Gahbauer, F., et al.\ 2008, \apj, 678, 262

\bibitem[Bartoli et al.(2014)]{2014ChPhC..38d5001B} Bartoli, B., Bernardini, P., J.~Bi, X., et al.\ 2014, Chinese Physics C, 38, 045001

\bibitem[Lewis \& Bridle(2002)]{2002PhRvD..66j3511L} Lewis, A., \& Bridle, S.\ 2002, \prd, 66, 103511

\bibitem[Liu et al.(2012)]{2012PhRvD..85d3507L} Liu, J., Yuan, Q., Bi, X.-J., Li, H., \& Zhang, X.\ 2012, \prd, 85, 043507

\bibitem[Amenomori et al.(2008)]{2008ApJ...678.1165A} Amenomori, M., Bi, X.~J., Chen, D., et al.\ 2008, \apj, 678, 1165

\bibitem[Ogio et al.(2004)]{2004ApJ...612..268O} Ogio, S., Kakimoto, F., Kurashina, Y., et al.\ 2004, \apj, 612, 268

\bibitem[Yoon et al.(2011)]{2011ApJ...728..122Y} Yoon, Y.~S., Ahn, H.~S., Allison, P.~S., et al.\ 2011, \apj, 728, 122

\bibitem[Bartoli et al.(2015)]{2015arXiv150203164B} Bartoli, B., Bernardini, P., Bi, X.~J., et al.\ 2015, arXiv:1502.03164

\bibitem[Blandford \& Eichler(1987)]{1987PhR...154....1B} Blandford, R., \& Eichler, D.\ 1987, \physrep, 154, 1

\bibitem[Cesarsky(1980)]{1980ARA&A..18..289C} Cesarsky, C.~J.\ 1980, \araa, 18, 289

\bibitem[Letaw et al.(1993)]{1993ApJ...414..601L} Letaw, J.~R., Silberberg, R., \& Tsao, C.~H.\ 1993, \apj, 414, 601

\bibitem[Ahn et al.(2009)]{2009ApJ...707..593A} Ahn, H.~S., Allison, P., Bagliesi, M.~G., et al.\ 2009, \apj, 707, 593

\bibitem[Bartoli et al.(2012)]{2012PhRvD..85i2005B} Bartoli, B., Bernardini, P., Bi, X.~J., et al.\ 2012, \prd, 85, 092005

\bibitem[Shibata et al.(2010)]{2010ApJ...716.1076S} Shibata, M., Katayose, Y., Huang, J., \& Chen, D.\ 2010, \apj, 716, 1076

\bibitem[Antoni et al.(2005)]{2005APh....24....1A} Antoni, T., Apel, W.~D., Badea, A.~F., et al.\ 2005, Astroparticle Physics, 24, 1

\bibitem[Apanasenko et al.(2001)]{2001APh....16...13A} Apanasenko, A.~V., Sukhadolskaya, V.~A., Derbina, V.~A., et al.\ 2001, Astroparticle Physics, 16, 13

\bibitem[Asakimori et al.(1998)]{1998ApJ...502..278A} Asakimori, K., Burnett, T.~H., Cherry, M.~L., et al.\ 1998, \apj, 502, 278

\bibitem[Glasmacher et al.(1999)]{1999APh....10..291G} Glasmacher, M.~A.~K., Catanese, M.~A., Chantell, M.~C., et al.\ 1999, Astroparticle Physics, 10, 291

\bibitem[Fowler et al.(2001)]{2001APh....15...49F} Fowler, J.~W., Fortson, L.~F., Jui, C.~C.~H., et al.\ 2001, Astroparticle Physics, 15, 49

\bibitem[H{\"o}randel(2001)]{2001ICRC....1...71H} H{\"o}randel, J.~R.\ 2001, International Cosmic Ray Conference, 1, 71

\bibitem[Bird et al.(1993)]{1993PhRvL..71.3401B} Bird, D.~J., Corbat{\'o}, S.~C., Dai, H.~Y., et al.\ 1993, Physical Review Letters, 71, 3401

\bibitem[H{\"o}randel(2004)]{2004APh....21..241H} H{\"o}randel, J.~R.\ 2004, Astroparticle Physics, 21, 241

\bibitem[Malkov \& O'C Drury(2001)]{2001RPPh...64..429M} Malkov, M.~A., \& O'C Drury, L.\ 2001, Reports on Progress in Physics, 64, 429

\bibitem[Reynolds(2008)]{2008ARA&A..46...89R} Reynolds, S.~P.\ 2008, \araa, 46, 89


\end{thebibliography}
\end{document}